\documentclass[letterpaper,conference]{IEEEtran}
\usepackage{cite}
\usepackage{amsmath,amssymb,amsfonts}
\usepackage{algorithmic}
\usepackage{graphicx}
\usepackage{textcomp}
\usepackage[table]{xcolor}
\usepackage[caption=false]{subfig}
\usepackage{amsmath}
\usepackage{multirow}
\usepackage{soul}
\usepackage[]{nohyperref}  % This makes hyperref commands do nothing without errors
\usepackage{url}  % This makes \url work
\def\BibTeX{{\rm B\kern-.05em{\sc i\kern-.025em b}\kern-.08em
	T\kern-.1667em\lower.7ex\hbox{E}\kern-.125emX}}
\begin{document}

\title{Lessons Learned: A Smart Campus Environment Using LoRaWAN}
	\author{\IEEEauthorblockN{Hari Prabhat Gupta}
   	Dept. of CSE, IIT (BHU) Varanasi, India\\
	\IEEEauthorblockA{\{hariprabhat.cse\}@iitbhu.ac.in}
	}

\maketitle

\begin{abstract}
The deployment of LoRaWAN (Long Range Wide Area Network) in dynamic environments, such as smart campuses, presents significant challenges in optimizing network parameters like spreading factor (SF), transmission power (TxPower), and managing mobility while ensuring reliable communication. In this paper, we first introduce the fundamental concepts of short-range and long-range communication protocols, emphasizing the specific requirements and advantages of LoRaWAN in various applications. Next, we discuss smart space solutions that integrate Edge, Fog, and Cloud computing, illustrating how these paradigms work in conjunction with both short-range and long-range communication protocols to enhance data processing and decision-making capabilities in real time. We then present our insights and lessons learned from the deployment of LoRaWAN across the campus~\cite{conf/networking/PandeyKGRR24}, focusing on the challenges encountered and the strategies employed to address them. This work provides a comprehensive overview of the methodologies applied, the results achieved, and the implications for future research and practical applications in IoT-enabled smart environments.
\end{abstract}

\begin{IEEEkeywords}
LoRaWAN, smart campuses, communication protocols, network optimization, spreading factor, transmission power, Edge computing, Fog computing, Cloud computing, IoT, mobility management, signal propagation, path loss, smart space solutions.
\end{IEEEkeywords}

\section{Introduction}

The rapid advancement of Internet of Things (IoT) technologies has led to the emergence of smart campuses, where a wide array of devices and systems work together to enhance operational efficiency and improve user experience~\cite{7123563, conf/wcnc/0001G24,journals/tpds/MishraGBD24}. In this context, the deployment of Long Range Wide Area Network (LoRaWAN) technology plays a pivotal role, providing low-power, long-range communication capabilities suitable for diverse applications ranging from environmental monitoring to asset tracking. However, deploying LoRaWAN in dynamic environments such as smart campuses presents significant challenges, particularly in optimizing network parameters like spreading factor (SF), transmission power (TxPower), and managing mobility, all while ensuring reliable communication.

This paper begins by introducing the fundamental concepts of short-range and long-range communication protocols, highlighting the unique features and requirements of LoRaWAN. We underscore the importance of these protocols in facilitating seamless connectivity among various devices within smart environments~\cite{9164991,5558084,6054047}. Next, we delve into smart space solutions that leverage Edge~\cite{7488250}, Fog, and Cloud computing paradigms~\cite{7123563,8016573,7498684}. These solutions enable efficient data processing and real-time decision-making by integrating both short-range and long-range communication protocols, thus enhancing the overall functionality of smart campus applications.

Subsequently, we present our findings and lessons learned from the deployment of LoRaWAN in a campus setting. This discussion encompasses the challenges encountered during deployment, such as signal propagation issues, network configuration complexities, and the effects of environmental factors on communication performance. We also explore strategies employed to address these challenges, offering insights into effective network management in dynamic environments.

The aim of this work is to provide a comprehensive overview of the methodologies applied in the deployment of LoRaWAN, the results achieved, and the implications for future research and practical applications. By sharing our experiences and findings, we hope to contribute valuable knowledge to the field of IoT and smart campus development, ultimately aiding in the optimization of network performance and reliability in similar dynamic environments~\cite{7123563,9130098,journals/csur/MishraG23,journals/wpc/ChopadeGD23,conf/iccps/0012G023,conf/infocom/MishraGD22}.

\section{Wireless Networking Protocol}
\begin{figure}[h]
    \centering
    \includegraphics[width=1\linewidth]{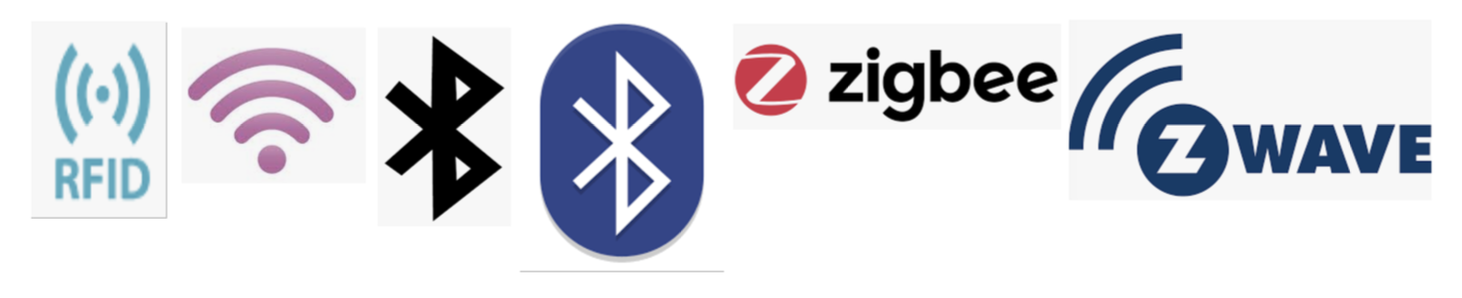}\\
    (a) Short-range communication protocols.\\
    \includegraphics[width=1\linewidth]{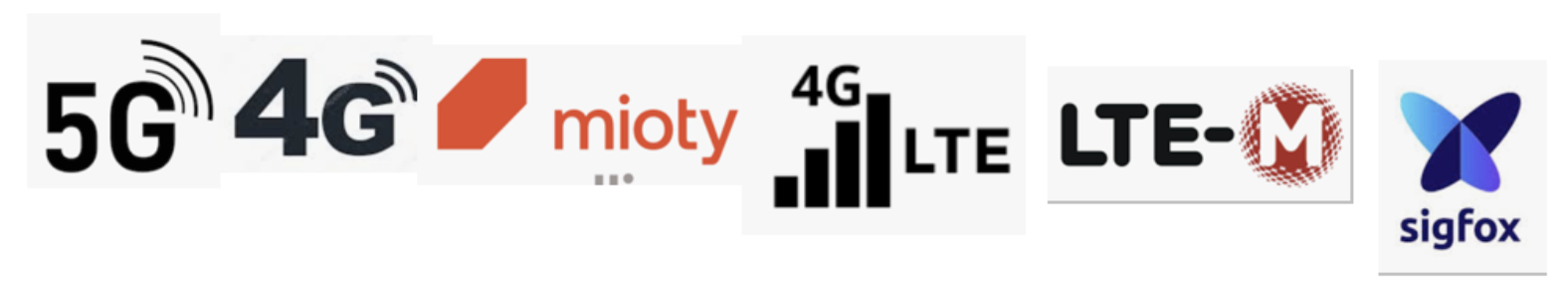}\\
    (b) Long-range communication protocols.
    \caption{Illustrates of short-range and long-range communication protocols.}
    \label{fig:enter-label}
\end{figure}

Wireless networking protocols play a crucial role in enabling communication between devices across various applications, including the Internet of Things (IoT). These protocols can be broadly classified into two main categories: short-range and long-range communication protocols, depending on their operational range, power requirements, and data rate capabilities. This section provides an overview of these categories and highlights several widely used protocols within each type. Figure~\ref{fig:enter-label} illustrates common short-range and long-range communication protocols.

\subsection{Short-range Communication Protocols}
Short-range communication protocols are designed to support high data rates and low power consumption over shorter distances, making them suitable for various applications. RFID (Radio Frequency Identification)~\cite{1589116,7123563} can operate at distances of up to 100 meters (for active tags) and is widely used in inventory management and asset tracking due to its very low power consumption. NFC (Near Field Communication)~\cite{9311219} operates at a range of up to 10 centimeters, enabling contactless payments and access control while maintaining very low power usage. Zigbee~\cite{4460126}, another low-power protocol, supports a range of 10 to 100 meters with a data rate of 250 kbps, making it ideal for home automation and wireless sensor applications. Thread offers a range of 10 to 100 meters and a data rate of 250 kbps, focusing on smart home devices. Z-Wave~\cite{7745306} operates at a similar range of up to 100 meters, with data rates between 9.6 to 100 kbps, catering to home automation and smart devices. Bluetooth supports a range of up to 100 meters with data rates of up to 3 Mbps and is commonly found in personal area networks and wearable technology. Wi-Fi~\cite{conf/wcnc/ThangadoraiSGK24}, known for its high data rates (up to 1 Gbps) and range (50-100 meters indoors), requires higher power consumption, making it suitable for high-data IoT applications and internet access. Lastly, Infrared (IR) communication, operating at up to 5 meters with data rates up to 4 Mbps, is primarily used for remote controls and data transfer between devices. M-Bus (Meter-Bus), used for utility metering, supports a range of up to 1 kilometer and varies in data rate while offering low power consumption. Collectively, these protocols enable a wide range of applications, from smart home automation to industrial IoT solutions, highlighting the diversity and versatility of short-range communication technologies. Table~\ref{tab:short_range} illustrates the trade-offs for Long-range communication protocols.

\begin{table}[t]
\centering
\caption{Trade-offs for Short-range Communication Protocols}
\begin{tabular}{|p{1.5cm}|p{1cm}|p{1cm}|p{1cm}|p{1.5cm}|}
\hline
\textbf{Protocol} & \textbf{Range} & \textbf{Data Rate} & \textbf{Power Consumption} & \textbf{Typical Applications} \\ \hline
\textbf{RFID (Radio Frequency Identification)} & Up to 100 m (active) & Varies (low) & Very Low & Inventory Management, Asset Tracking \\ \hline
\textbf{NFC (Near Field Communication)} & Up to 10 cm & Up to 424 kbps & Very Low & Contactless Payments, Access Control \\ \hline
\textbf{Zigbee} & 10-100 m & 250 kbps & Very Low & Home Automation, Wireless Sensors \\ \hline
\textbf{Infrared (IR)} & Up to 5 m & Up to 4 Mbps & Low & Remote Controls, Data Transfer Between Devices \\ \hline
\textbf{M-Bus (Meter-Bus)} & Up to 1 km & Varies & Low & Utility Metering, Smart Metering \\ \hline
\textbf{Thread} & 10-100 m & 250 kbps & Low & Smart Home Devices, IoT Applications \\ \hline
\textbf{Z-Wave} & Up to 100 m & 9.6 - 100 kbps & Low & Home Automation, Smart Home Devices \\ \hline
\textbf{Bluetooth} & Up to 100 m & Up to 3 Mbps & Low to Medium & Personal Area Networks, Wearables \\ \hline
\textbf{Wi-Fi} & 50-100 m (indoor) & Up to 1 Gbps & High & Internet Access, High Data IoT \\ \hline
\end{tabular}
\label{tab:short_range}
\end{table}

\subsection{Long-range Communication Protocols}
Long-range communication protocols are essential for applications requiring connectivity over extended distances, often in outdoor or expansive environments. These protocols are designed to provide reliable communication while balancing data rate, range, and power consumption. Sigfox~\cite{8480255} is a popular long-range protocol, supporting a range of up to 50 kilometers in rural areas, with low data rates typically around 100 bps, suited for IoT devices that transmit small amounts of data infrequently. NB-IoT (Narrowband IoT~\cite{8170296,8502812}) operates over existing cellular networks and can cover distances up to 35 kilometers, providing higher data rates compared to Sigfox while maintaining low power consumption, making it suitable for smart cities and utility metering. LTE-M (Long Term Evolution for Machines~\cite{7880946}) is also designed for cellular IoT, offering higher data rates (up to 1 Mbps) and lower latency over a similar range as NB-IoT, making it suitable for applications requiring more data throughput, such as connected vehicles and wearables. 4G (Fourth Generation) and 5G (Fifth Generation) networks offer extensive coverage and high data rates, with 4G capable of supporting speeds of up to 1 Gbps and 5G potentially reaching 10 Gbps, making them suitable for a wide range of applications, from mobile broadband to critical IoT applications. Mioty~\cite{10176016}, a relatively new protocol, excels in connecting numerous devices over long distances, supporting ranges of up to 10 kilometers with low power consumption, making it ideal for industrial IoT and smart city applications. Satellite communication, although generally more expensive, provides global coverage and is crucial for remote monitoring applications in agriculture, forestry, and environmental management. Overall, these long-range communication protocols enable various applications that require reliable connectivity over significant distances, catering to the growing demands of IoT and smart technologies. Table~\ref{tab:long_range} illustrates the trade-offs for long-range communication protocols.
\begin{table}[t]
\centering
\caption{Trade-offs for Long-range Communication Protocols}
\begin{tabular}{|p{1.5cm}|p{1cm}|p{1cm}|p{1cm}|p{1.5cm}|}
\hline
\textbf{Protocol} & \textbf{Range} & \textbf{Data Rate} & \textbf{Power Consumption} & \textbf{Typical Applications} \\ \hline
\textbf{Sigfox} & Up to 50 km & 100 bps & Very Low & IoT Devices, Asset Tracking \\ \hline
\textbf{NB-IoT (Narrowband IoT)} & Up to 35 km & Up to 250 kbps & Low & Smart Cities, Utility Metering \\ \hline
\textbf{LTE-M (Long Term Evolution for Machines)} & Up to 35 km & Up to 1 Mbps & Low to Medium & Connected Vehicles, Wearables \\ \hline
\textbf{4G (Fourth Generation)} & Up to 30 km & Up to 1 Gbps & Medium to High & Mobile Broadband, IoT Applications \\ \hline
\textbf{5G (Fifth Generation)} & Up to 10 km & Up to 10 Gbps & Medium to High & Enhanced Mobile Broadband, Critical IoT \\ \hline
\textbf{Mioty} & Up to 10 km & Up to 1 Mbps & Low & Industrial IoT, Smart Cities \\ \hline
\textbf{Satellite Communication} & Global Coverage & Varies & High & Remote Monitoring, Agriculture \\ \hline
\end{tabular}
\label{tab:long_range}
\end{table}
\subsection{Challenges in short-range and long-range Communication Protocols}
Short-range communication protocols are widely used for various applications, particularly in personal area networks and smart home devices. The main challenge for these protocols is their limited range, which typically spans from a few centimeters to about a hundred meters, depending on the protocol. This limited distance can be a significant constraint in larger environments, such as warehouses or outdoor settings, where multiple devices need to communicate over extended areas. Additionally, physical obstacles like walls, furniture, and other barriers can degrade signal strength, leading to interruptions in communication. Interference from other wireless devices operating in the same frequency bands, especially in crowded urban environments, can further exacerbate these issues, making it challenging to maintain stable connections. On the other hand, one of the key advantages of short-range protocols is their energy consumption. Many of these protocols are designed to be energy-efficient, allowing devices to operate effectively for extended periods without frequent battery replacements or recharging, which is essential for applications in smart home systems that utilize numerous sensors and devices.

Long-range communication protocols, including technologies like Sigfox, NB-IoT, and LTE-M, are designed to facilitate connectivity over extended distances, often in outdoor or rural environments. The main challenge in long-range communication is energy consumption, as devices need to maintain reliable communication over several kilometers, which can require more power than their short-range counterparts. While these protocols can transmit data over long distances, they typically do so at lower data rates. For instance, Sigfox operates at around 100 bps, which is suitable for applications that require infrequent transmission of small data packets but may not be adequate for real-time data needs. Moreover, the infrastructure requirements for these protocols can be substantial. Cellular-based systems, such as 4G and 5G, necessitate the establishment of extensive cellular networks, which can be costly and time-consuming to deploy and maintain, especially in remote areas. Despite these challenges, a significant advantage of long-range communication protocols is their ability to connect devices over vast distances, making them essential for applications like smart city initiatives, agricultural monitoring, and remote asset tracking, where connectivity across expansive areas is crucial.

\section{The LoRa Protocol: A Solution for Long-Range Communication with Minimal Energy}
\begin{figure}[h]
    \centering
    \includegraphics[width=1\linewidth]{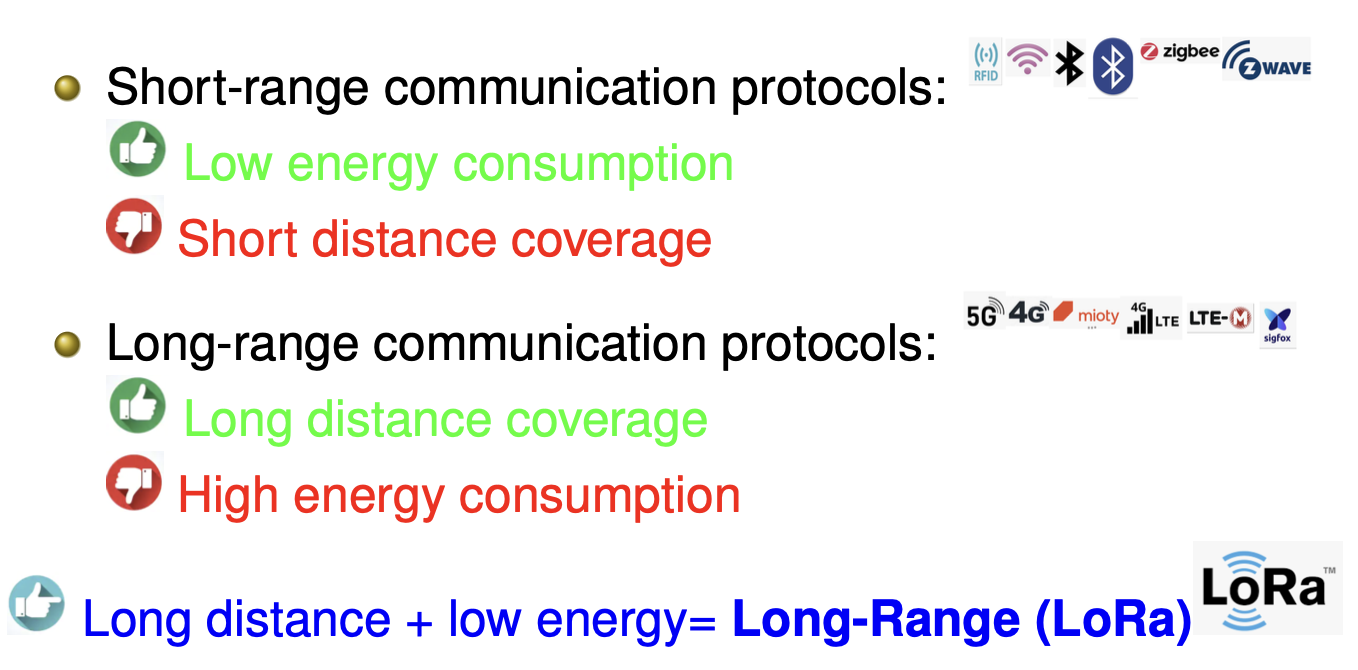}
    \caption{Illustrates the advantages and limitations of the protocols.}
    \label{adv_shoer_long}
\end{figure}

Among the various long-range communication protocols, LoRa (Long Range) stands out as a particularly effective solution to many of these challenges~\cite{conf/wcnc/PandeyKG024,conf/icc/KumariGDB23}. LoRa technology is specifically designed for low-power, wide-area networks (LPWAN), enabling devices to communicate over distances of up to 15 kilometers while consuming minimal energy. Its unique chirp spread spectrum modulation technique provides robust resistance to interference, making it ideal for deployment in urban and rural environments alike. This capability allows for reliable data transmission even in challenging conditions, such as areas with high signal congestion or variable terrain. LoRa’s low power consumption addresses the energy consumption challenge prevalent in other long-range protocols, enabling devices to operate for years on small batteries~\cite{conf/wowmom/KumariGDD22,7815384,conf/mswim/KumariGM021}. This is crucial for applications like smart agriculture and environmental monitoring, where devices may be deployed in remote locations without easy access to power sources. Figure~\ref{adv_shoer_long} illustrates the advantages and limitations of the communication protocols.

Furthermore, LoRa's architecture supports a large number of connected devices within a single network, making it an excellent choice for IoT applications that require extensive sensor deployments. Its scalability and flexibility allow for the integration of various devices and applications, fostering the development of smart city initiatives, industrial automation, and environmental monitoring solutions. The ability to transmit small data packets at low power while maintaining long-range connectivity makes LoRa particularly appealing for scenarios that demand periodic data updates, such as temperature or humidity monitoring in agricultural fields. Overall, LoRa effectively addresses the limitations faced by traditional long-range communication protocols, offering a reliable and efficient solution that meets the evolving needs of modern IoT applications.

\subsection{About LoRa}
Long Range (LoRa) is a low-power, wide-area networking protocol designed for IoT applications. It enables long-range communication by utilizing chirp spread spectrum technology, which allows for robust transmission over significant distances while consuming minimal power. LoRa is particularly suited for applications where battery life is crucial and where devices need to operate in challenging environments. Its ability to penetrate dense urban structures makes it an ideal choice for smart city applications and remote monitoring systems.

\subsection{Applications of LoRa}
\begin{figure}[h]
    \centering
    \includegraphics[width=1\linewidth]{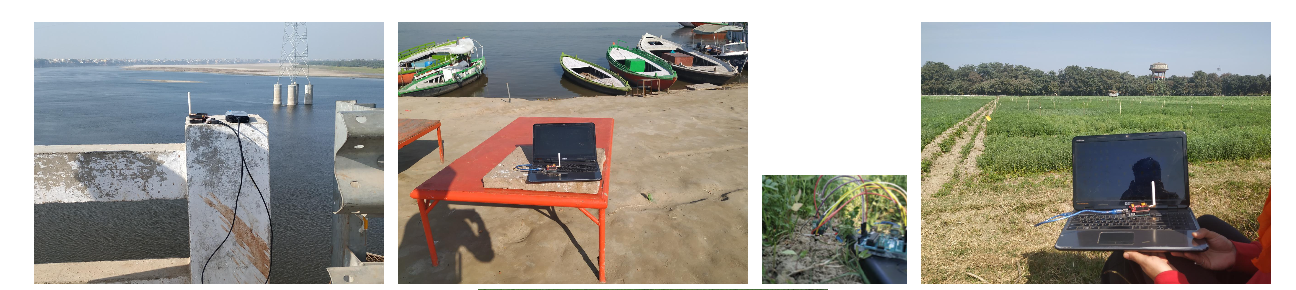}\\
    \includegraphics[width=1\linewidth]{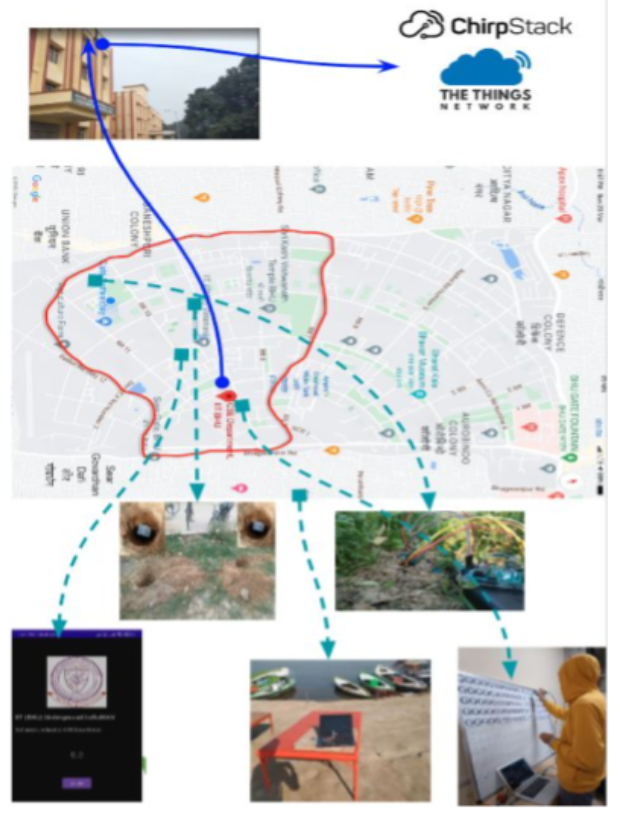}
    \caption{Illustration of applications of LoRa. The deplyment done at IIT (BHU) Varanasi.}
    \label{LoRa_application}
\end{figure}

LoRa technology has gained prominence across multiple industries due to its long-range communication capabilities combined with low power consumption~\cite{conf/wowmom/MishraKGSDSP21,7803607,conf/infocom/KumariGD21,7377400,8326735,conf/icc/KumariGD20,journals/cem/GhoshMGSR22}. One key application is water pollution monitoring, where sensors deployed in rivers, lakes, and other water bodies continuously collect data on pollution levels, transmitting it over long distances to central servers for real-time analysis~\ref{LoRa_application}. This allows authorities to take swift corrective actions to protect water quality. In agriculture, LoRa is indispensable for precision farming, enabling farmers to monitor soil moisture, weather conditions, and crop health in vast fields. LoRa’s ability to cover extensive areas with minimal energy consumption makes it ideal for agricultural environments where power resources are often limited. Another important application is road health monitoring, where sensors installed along roadways track traffic conditions, structural integrity, and surface quality. These sensors can transmit data to central servers even in remote areas, providing valuable insights for maintaining and upgrading infrastructure.

LoRa stands out from other communication protocols such as Wi-Fi or cellular networks (like 4G or 5G) due to its ability to cover long distances with minimal energy usage. Wi-Fi typically offers high data rates but is limited by its short range, making it unsuitable for expansive, rural, or remote applications. Cellular networks, while offering broader coverage, often come with higher energy demands and infrastructure costs, which are not ideal for low-power IoT devices deployed in large quantities. LoRa’s chirp spread spectrum modulation allows it to handle interference better than protocols like Zigbee, making it more reliable in dense or noisy environments. Thus, for applications requiring long-range, low-power, and cost-effective communication, especially in remote or hard-to-reach locations, LoRa offers an unmatched balance of coverage, energy efficiency, and scalability.

\begin{figure*}[t]
    \centering
    \includegraphics[width=1\linewidth]{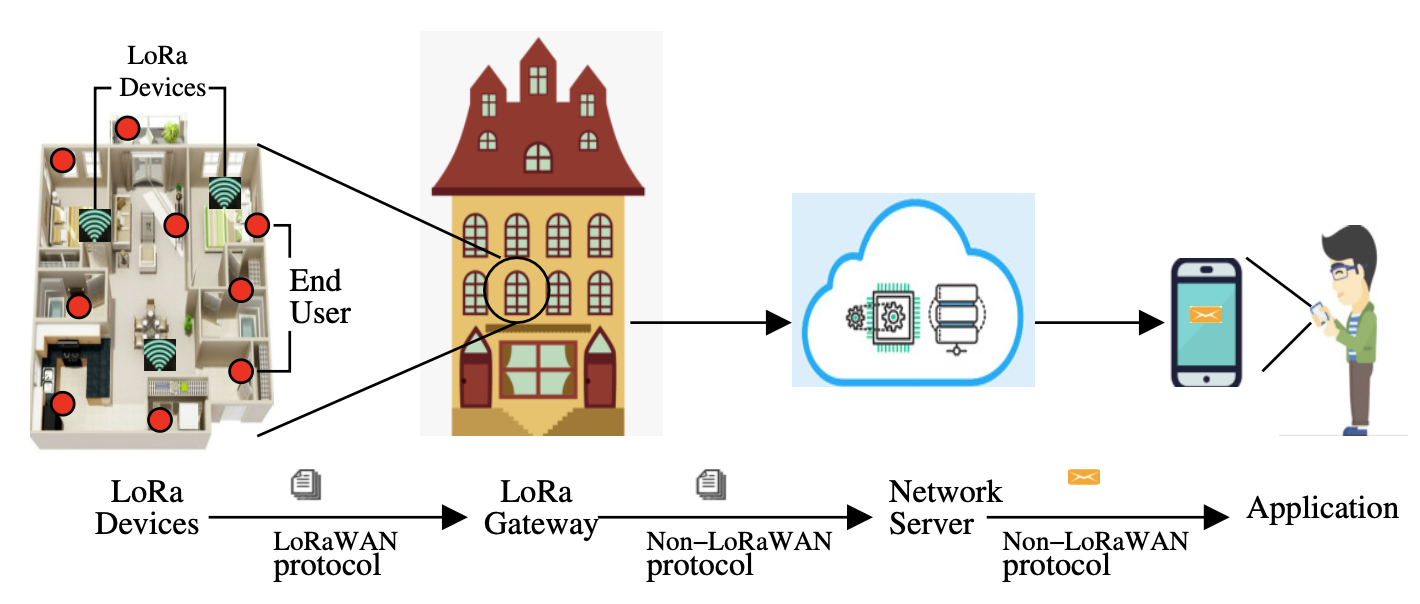}
    \caption{Illustrates the architecture of LoRa}
    \label{fig:architecture}
\end{figure*}
\subsection{Architecture of LoRa}
LoRa technology is structured in a way that enables efficient, long-range communication between sensors and end users while maintaining low power consumption. This architecture is comprised of several key components that work together to transmit, process, and analyze data collected from the environment. Figure~\ref{fig:architecture} illustrates the architecture of LoRa with different components.

First, sensors play a crucial role in the system. These devices are responsible for collecting environmental data such as temperature, humidity, air quality, or water levels, depending on the application. These sensors can be placed in various locations, continuously monitoring the environment and transmitting their readings to the next stage of the network.

The LoRa Node (LN) is the endpoint device that interfaces directly with the sensors. It receives data from the sensors, processes it, and then uses the LoRa communication protocol to send the data over long distances to the LoRa Gateway (LG). The LoRa node is energy-efficient, allowing it to operate for extended periods in remote or hard-to-access areas.

The LoRa Gateway (LG) is the intermediary device that collects data from multiple LoRa nodes in its range. It aggregates this data and forwards it to a central Network Server (NS). The gateway serves as a bridge between the low-power, long-range LoRa nodes and the higher-level infrastructure, such as the Cloud or local servers, where more advanced processing can take place.

Once the data reaches the Network Server (NS), it manages the entire network's communication and ensures that data packets are correctly routed. The network server performs tasks such as error checking, managing device registrations, and forwarding the data to the Application Server (AS), where the real-time data analytics and actionable insights are generated.

Finally, the End User (EU) is the person or system that interacts with the processed data, which is typically presented through a user-friendly interface. Based on this data, users can make informed decisions, such as adjusting irrigation in a smart agriculture system or responding to environmental conditions in a smart city application. This architecture's modular design allows for flexible deployment across a range of industries, enabling the efficient transmission of sensor data over long distances with minimal energy consumption.

\subsection{Properties of LoRa}
LoRa technology possesses several key properties that make it suitable for long-range, low-power communication applications. One of the fundamental aspects of LoRa is its star topology, where all sensor nodes communicate with a central gateway. This centralized design simplifies the network's overall architecture and ensures efficient management of data transmission. The gateway serves as the primary point for receiving and forwarding data from multiple nodes, reducing the complexity and cost of deploying large networks in applications like smart agriculture or environmental monitoring.

LoRa operates in the unlicensed Industrial, Scientific, and Medical (ISM) bands, which is another notable advantage. The use of unlicensed spectrum allows for broader and easier deployment across different regions without requiring users to obtain specific regulatory approval. This makes LoRa highly scalable and accessible for a variety of applications in industries such as smart cities and infrastructure monitoring.

The duty cycle constraint is a regulatory requirement that limits the amount of time a LoRa device can spend transmitting data on the shared ISM band. This ensures fair access to the spectrum and prevents any single device from monopolizing the communication channel. While this constraint limits the total transmission time, LoRa’s low data rate and energy efficiency compensate for this, making it ideal for applications where infrequent data transmission is sufficient, such as utility metering and environmental sensing.

Another essential feature of LoRa is its bi-directional communication capability, allowing devices not only to send data to the server but also to receive control commands and acknowledgments. This two-way communication makes LoRa suitable for applications requiring real-time feedback or control, such as remote monitoring systems where users can adjust parameters or receive notifications based on sensor data.

Finally, LoRa employs Chirp Spread Spectrum (CSS) modulation, which enhances the signal’s robustness against noise and interference. CSS modulation spreads the signal over a wide frequency range, making it more resistant to interference and allowing for reliable communication even in challenging environments. This ensures that LoRa can maintain a strong, stable connection in areas with high levels of radio frequency interference or in urban environments with numerous other wireless devices operating simultaneously. 

\subsection{Resources of LoRa}
The resources of LoRa technology play a crucial role in determining its performance across different applications. The carrier frequency in LoRaWAN deployment is flexible, starting with a minimal setup of 8 channels, which can be expanded to 16 channels to accommodate a larger number of devices and reduce interference in densely populated networks. This allows for scalability in various IoT applications, where multiple devices need to communicate efficiently.

The bandwidth of the transmission channel significantly influences data rates and coverage. In LoRa, a bandwidth of 125 kHz corresponds to a chip rate of 125 kcps, striking a balance between data throughput and the range of communication. Narrower bandwidths extend the communication range but may reduce data transfer speed, making this a key consideration when designing LoRa networks for long-range yet low-data-rate applications.

Transmission power in LoRa is adjustable, ranging from 2 dBm to 20 dBm. This flexibility allows optimization based on the specific requirements of the application, such as the distance to the gateway or energy consumption limitations. For example, in remote sensing applications, lower power settings may be sufficient, conserving battery life while maintaining reliable communication.

Finally, LoRa uses coding rate as part of its forward error correction (FEC) mechanism, which provides resilience against interference. The coding rate can be configured between 4/5, 4/6, 4/7, and 4/8, allowing network designers to choose between higher reliability (with more error correction) and better throughput. This adaptability is vital for ensuring data integrity in challenging environments, where signal interference might otherwise degrade communication.

\subsection{Challenges and Overview of Solutions in LoRa}
LoRa technology offers significant advantages for long-range, low-power communication, but it also faces several challenges that must be managed to fully exploit its potential. One key issue is limited resources, as LoRa typically supports only six virtual channels (SF6-SF12). In high-density environments with numerous devices, this limitation can lead to network congestion, impacting performance. Efficient channel allocation and network management strategies are required to mitigate these issues and ensure smooth operation in large-scale deployments.

Another challenge is interference, particularly co-SF (same spreading factor) and inter-SF interference, which can degrade communication quality. This interference is common in unlicensed frequency bands where multiple devices share the same spectrum. Developing advanced interference management techniques and improving spectrum efficiency are critical for enhancing LoRa's performance, especially in urban or industrial environments with many devices. Additionally, LoRa's low data rate, capped at 27 kbps, can be a constraint for applications requiring higher throughput, such as real-time video streaming or large data transfers. Compared to faster technologies like Bluetooth Low Energy (BLE), this may limit its use in certain data-intensive scenarios.

Moreover, data transmission issues arise from operating in crowded license-free frequency bands (such as the 865 MHz – 867 MHz range in India), where increased traffic can lead to data collisions and packet loss. LoRa also operates under strict duty cycle constraints, limiting the frequency of transmissions to comply with regulations, which can impact applications needing continuous or real-time monitoring. Lastly, security in LoRa networks is another concern, as the lightweight Edge devices must implement effective encryption and authentication mechanisms without compromising energy efficiency or performance. Addressing these challenges is vital for unlocking the full potential of LoRa, enabling its wide application across smart cities, agriculture, environmental monitoring, and more.

\section{Smart Space System using LoRaWAN}
Smart spaces utilize both short-range and long-range communication protocols to effectively collect sensory data from various locations within the environment. These protocols facilitate the monitoring of human activity and other interactions occurring within the space~\cite{6780609,9166711,5567086,9130098,7460727,9479778,1321026,9164991,5558084,6054047}. The architecture of a smart space typically involves Edge devices~\cite{conf/sensys/KumariG023}, Fog devices, and Cloud computing, each playing a crucial role in data processing and communication. In the previous section, we discussed the characteristics of short-range and long-range communication protocols. 

In this section, we will first examine the functionality of Edge devices and Fog devices, highlighting their roles in the smart space ecosystem. We will also explore the necessity of using both short-range and long-range communication methods in these devices. Furthermore, this section will address the challenges encountered when transitioning a conventional space into a smart space, emphasizing the technical and logistical issues that must be overcome to achieve a fully functional smart environment.

\subsection{Smart Space System}
\begin{figure*}
    \centering
    \includegraphics[width=1\linewidth]{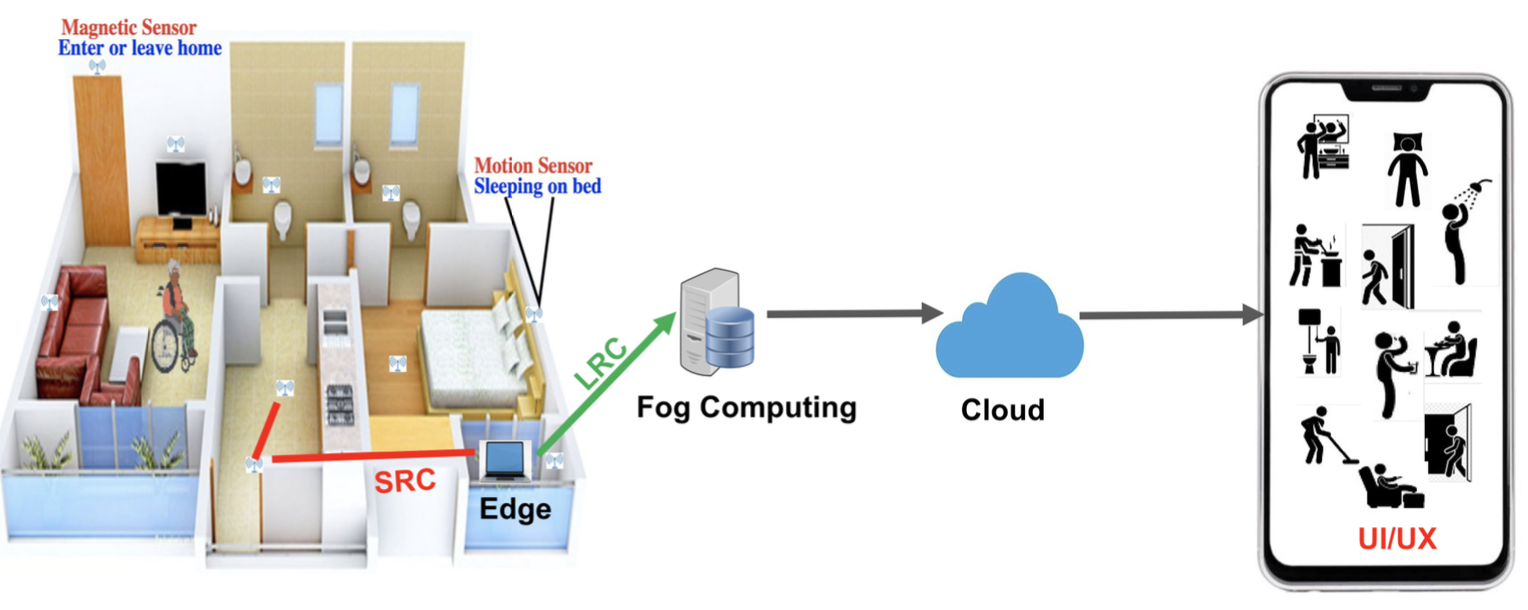}
    \caption{Illustration of a smart space system.}
    \label{fig:smartspace}
\end{figure*}

An intelligent environment equipped with interconnected sensors and communication technologies can be scaled as a smart space. The system considers the interaction between Edge, Fog devices, and various sensors, facilitating seamless data acquisition, communication, and processing across multiple layers. Figure~\ref{fig:smartspace} illustrates a smart space system, showcasing the integration of these components to enable real-time decision-making, efficient resource management, and enhanced connectivity.

In a smart space system such as smart home environment, Edge devices play a critical role in gathering data from various sensors deployed throughout the space. These sensors monitor key environmental factors such as temperature, humidity, lighting, and occupancy levels, as well as other parameters like air quality and energy consumption. For example, in a smart home, sensors installed in different rooms can track whether individuals are present, adjust lighting, or regulate the thermostat based on the number of occupants. Once the Edge devices receive data from these sensors, they aggregate the information and perform preliminary processing, such as removing noise from the data or converting it into a standardized format.

After processing, the Edge devices use Short-Range Communication (SRC) protocols, such as Zigbee, Bluetooth, or Wi-Fi, to transmit the data to local Fog devices. These Fog devices are located within the home and act as intermediaries between the Edge devices and the Cloud. They perform more advanced operations like filtering, data aggregation, and local analytics to reduce the volume of information sent to the Cloud. For instance, the Fog device might analyze patterns from occupancy sensors to predict the best times to automatically adjust heating or cooling systems, improving energy efficiency. By offloading some of the processing to Fog devices, the system can respond quickly without needing to rely entirely on Cloud resources.

In cases where further, more comprehensive analysis is needed, Fog devices utilize Long-Range Communication (LRC) protocols, such as LoRa or LTE, to send processed data to the Cloud. The Cloud servers handle more complex tasks like integrating data from multiple sources (e.g., other smart homes or public data) and generating predictive insights, such as detecting potential energy wastage patterns or alerting homeowners about maintenance issues. Once the Cloud has processed the data, the results are sent back to the Fog devices, which in turn relay the final insights to the end users, often through a mobile app or other interfaces. For example, the homeowner might receive a notification on their phone about energy savings, or the system might automatically adjust settings to optimize comfort and efficiency. This architecture, which leverages both short-range and long-range communication, ensures a balance between responsiveness and comprehensive data analysis while reducing the load on centralized Cloud systems.In a smart home environment, Edge devices play a critical role in gathering data from various sensors deployed throughout the space. These sensors monitor key environmental factors such as temperature, humidity, lighting, and occupancy levels, as well as other parameters like air quality and energy consumption. For example, in a smart home, sensors installed in different rooms can track whether individuals are present, adjust lighting, or regulate the thermostat based on the number of occupants. Once the Edge devices receive data from these sensors, they aggregate the information and perform preliminary processing, such as removing noise from the data or converting it into a standardized format.

After processing, the Edge devices use Short-Range Communication (SRC) protocols, such as Zigbee, Bluetooth, or Wi-Fi, to transmit the data to local Fog devices. These Fog devices are located within the home and act as intermediaries between the Edge devices and the Cloud. They perform more advanced operations like filtering, data aggregation, and local analytics to reduce the volume of information sent to the Cloud. For instance, the Fog device might analyze patterns from occupancy sensors to predict the best times to automatically adjust heating or cooling systems, improving energy efficiency. By offloading some of the processing to Fog devices, the system can respond quickly without needing to rely entirely on Cloud resources.

In cases where further, more comprehensive analysis is needed, Fog devices utilize Long-Range Communication (LRC) protocols, such as LoRa or LTE, to send processed data to the Cloud. The Cloud servers handle more complex tasks like integrating data from multiple sources (e.g., other smart homes or public data) and generating predictive insights, such as detecting potential energy wastage patterns or alerting homeowners about maintenance issues. Once the Cloud has processed the data, the results are sent back to the Fog devices, which in turn relay the final insights to the end users, often through a mobile app or other interfaces. For example, the homeowner might receive a notification on their phone about energy savings, or the system might automatically adjust settings to optimize comfort and efficiency. This architecture, which leverages both short-range and long-range communication, ensures a balance between responsiveness and comprehensive data analysis while reducing the load on centralized Cloud systems.

\subsection{Smart Space System using LoRaWAN}
The Long-Range Communication (LRC) protocol can be LoRaWAN due to its capability to support low-power, wide-area networks. LoRaWAN offers several advantages, such as energy efficiency, long-range communication (up to 15 km), scalability, and the ability to operate in unlicensed ISM bands. These features make it an ideal choice for smart space systems where a large number of sensors need to communicate over extended distances with minimal energy consumption. The deployment procedure for a LoRaWAN-based smart space system consists of the following steps:

\begin{itemize}
\item \textbf{Step 1: Deploy sensors in strategic locations throughout the smart space to monitor environmental conditions, occupancy, and other relevant data points.} 

The first step in setting up a smart space system involves deploying sensors across different areas of the space to gather data on various environmental and operational parameters. These sensors can monitor a wide range of variables such as temperature, humidity, light levels, air quality, motion, and occupancy. The specific placement of the sensors is critical to ensuring comprehensive coverage of the space and capturing accurate data. 

For instance, in a smart office building, sensors might be placed in rooms to detect human presence and automatically adjust lighting and HVAC systems based on occupancy. In a smart home, sensors could be deployed to monitor indoor air quality and control ventilation systems or to detect water leaks and send alerts in case of emergencies. Environmental sensors could also monitor factors such as noise levels or carbon dioxide concentration to ensure that the environment remains safe and comfortable.

When choosing the location for each sensor, it is important to account for both the physical layout of the space and the type of data to be collected. For example, temperature sensors should be placed away from direct sunlight or heat sources to avoid inaccurate readings, while motion sensors should cover areas with high traffic to detect movement effectively. Additionally, sensors need to be positioned in a way that ensures they are within the communication range of the LoRaWAN network, allowing them to transmit data efficiently to the LoRa-enabled Edge devices or directly to the LoRa gateway.

Proper sensor deployment is essential for the system’s overall performance, as well-placed sensors ensure that the data collected is reliable and provides meaningful insights into the space’s operations. By strategically placing sensors, the smart space can effectively monitor and manage its various environmental and functional aspects, leading to optimized energy use, enhanced comfort, and improved safety for its occupants.

   \item \textbf{Step 2: Use Short-Range Communication (SRC) protocols to collect the sensor data and transmit it to nearby LoRa-enabled Edge devices.}

After deploying sensors throughout the smart space, the next step is to ensure that the data collected by these sensors is transmitted to nearby Edge devices for initial processing. Short-Range Communication (SRC) protocols like Zigbee, Bluetooth, or Wi-Fi are commonly used for this purpose because they provide reliable, low-power communication over short distances. These protocols are particularly suited for transmitting data from sensors located within the same room or building as the Edge device, ensuring efficient data collection with minimal power consumption.

Once the sensor data is collected using these short-range protocols, it is transmitted to the nearby {LoRa-enabled Edge devices}. These Edge devices are capable of performing local processing on the sensor data, such as filtering, aggregation, or preliminary analysis, before forwarding the data to the LoRaWAN gateway for long-range communication. The use of short-range communication protocols in this stage minimizes energy consumption at the sensor level, as SRC protocols are designed for short-distance, low-power communication. This step ensures that the sensor network operates efficiently, conserving energy while maintaining reliable data transmission to the Edge devices.

\item \textbf{Step 3: Set up the LoRaWAN gateway to collect data from the Edge devices. The gateway serves as a bridge between the Edge devices and the central network server, facilitating long-range communication.}

After the sensor data has been transmitted to the Edge devices via Short-Range Communication (SRC) protocols, the next step is to set up the {LoRaWAN gateway}, a crucial component in enabling long-range, low-power communication. The gateway acts as an intermediary between the Edge devices that have aggregated sensor data and the central {network server}, which is responsible for managing and processing data across the smart space system.  

The LoRaWAN gateway is equipped with {long-range communication} capabilities, allowing it to receive data from multiple LoRa-enabled Edge devices spread across a wide geographic area. Unlike traditional communication systems that may require multiple access points or base stations for coverage, LoRaWAN gateways are designed to cover extensive distances—typically up to 15 kilometers in rural areas and several kilometers in urban environments, depending on the surrounding obstacles and network configurations. This long-range capability is one of the main advantages of LoRaWAN technology, as it reduces the number of gateways required to cover a large area, making it both cost-effective and scalable for smart space applications. Figure~\ref{fig:gateway} illustrates the placement of the Gateway at the top of the CSE department building, IIT (BHU) Varanasi.
\begin{figure}
    \centering
    \includegraphics[width=1\linewidth]{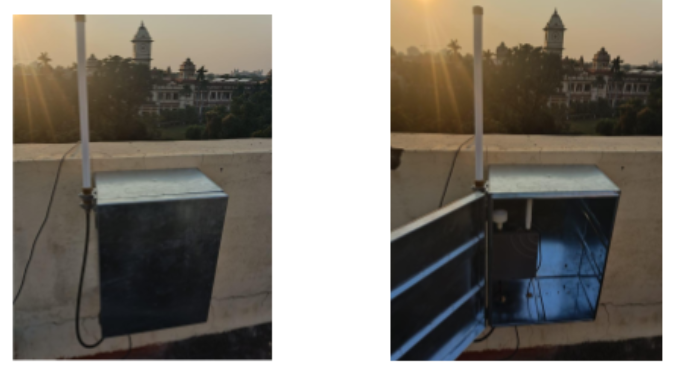}
    \caption{Illustration of the LoRaWNA Gateway at IIT (BHU) Varanasi.}
    \label{fig:gateway}
\end{figure}
The gateway receives data transmitted from the Edge devices using {Chirp Spread Spectrum (CSS) modulation}, a key feature of LoRa that ensures data can be transmitted reliably even in the presence of interference or over long distances. This modulation technique spreads the signal across a wide frequency range, improving its resistance to noise and allowing for robust communication even in environments with significant radio frequency interference, such as densely populated urban areas or industrial settings.

Once the gateway collects the data from the Edge devices, it forwards this information to the {network server} using a backhaul connection, which could be over the internet, cellular network, or even a wired Ethernet connection, depending on the deployment scenario. The gateway essentially translates the LoRa radio signals received from the Edge devices into IP-based packets that can be routed over standard networks to the central server. Figure~\ref{fig:server} illustrates the functionality of The Things Stack as a LoRaWAN Network Server.
\begin{figure}
    \centering
    \includegraphics[width=1\linewidth]{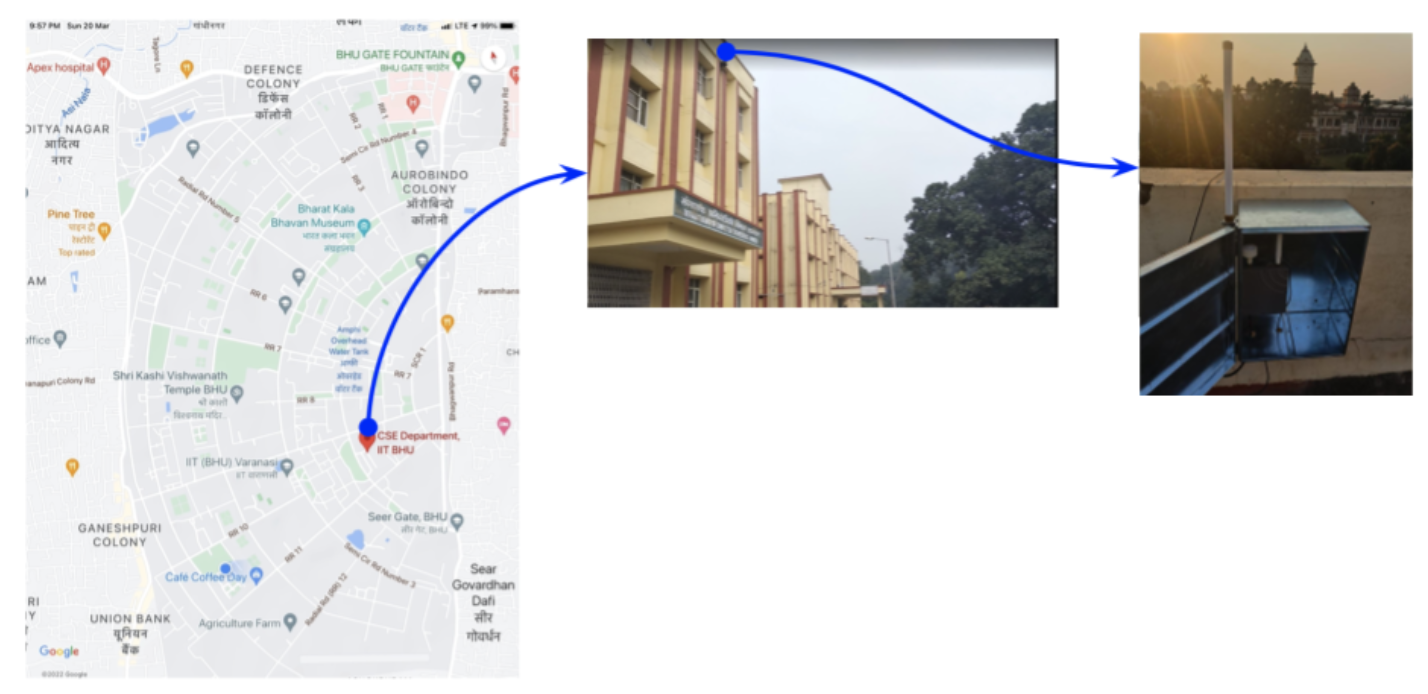}
    \includegraphics[width=1\linewidth]{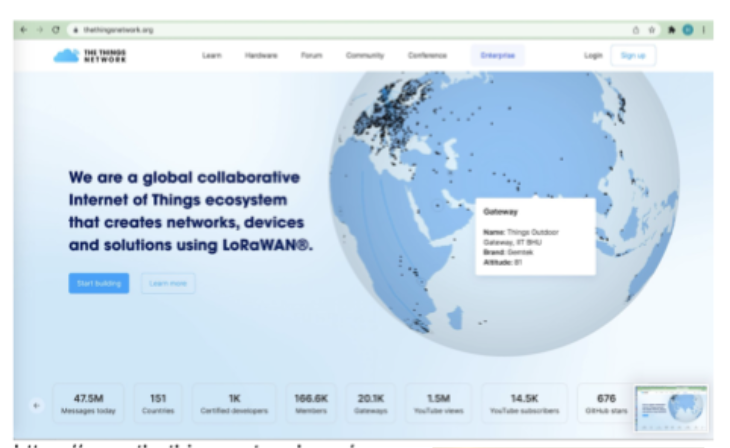}
\caption{Illustration of The Things Stack as a LoRaWAN Network Server}
    \label{fig:server}
\end{figure}

In addition to relaying data, the LoRaWAN gateway can also facilitate {bi-directional communication}. This means that not only can it send sensor data from the Edge devices to the network server, but it can also receive commands or updates from the network server and pass them back to the Edge devices. For instance, in a smart space environment, the network server might process the incoming sensor data and determine that certain actions are necessary, such as adjusting lighting levels, controlling HVAC systems, or sending alerts to users. These commands are transmitted back through the gateway to the relevant Edge devices, which then carry out the required actions.

To ensure optimal performance, multiple gateways can be deployed in a given area, providing {redundancy and improved coverage}. This multi-gateway setup is particularly useful in smart spaces where there are physical barriers or varying environmental conditions that may affect signal propagation. The network server is capable of receiving data from multiple gateways simultaneously, allowing it to select the best-quality transmission or aggregate the data received from different sources to improve accuracy and reliability.
\end{itemize}

\section{A Smart Campus Environment Using LoRaWAN}
In this section, we detail the insights gained through the extensive deployment of LoRaWAN technology as part of our smart campus initiative. The primary goal of this initiative was to leverage LoRaWAN's long-range, low-power communication capabilities to enable various smart applications such as environmental monitoring, smart agriculture, and campus-wide connectivity. Over the course of our deployment, we encountered several challenges that required deep technical analysis and innovative solutions~\cite{conf/networking/PandeyKGRR24}.

Our study addressed three critical aspects of LoRaWAN deployment: (1) configuring the network parameters to balance performance, energy efficiency, and coverage; (2) managing the transmission of large data packets over a network designed for low-data-rate communication; and (3) overcoming interference and connectivity issues that arose during the deployment of multiple LoRaWAN devices across different areas of the campus, which included a mix of open spaces, buildings, and other obstructions.

\subsection{Configuring LoRaWAN Network Parameters}

LoRaWAN operates with several key parameters such as Spreading Factor (SF), transmission power, and coding rate, each playing a critical role in determining the network’s overall performance. Configuring these parameters correctly is essential to achieve smooth connectivity, minimize delay, and optimize energy consumption. However, striking the right balance between these factors is challenging, especially since each parameter affects the network's performance in different ways. For example, a higher spreading factor increases communication range, but at the expense of increased transmission time and higher power consumption. Conversely, lowering the spreading factor conserves energy but reduces the communication range and robustness of the signal.

\subsubsection{The Challenge of Parameter Configuration}

Configuring network parameters becomes even more challenging when deploying LoRaWAN in an unknown or unvisited location. In such environments, site-specific factors such as terrain, building obstructions, and environmental conditions can drastically impact signal strength and connectivity. Typically, engineers are required to visit these sites to manually fine-tune the parameters for optimal performance. However, this manual process is both time-consuming and costly, particularly when dealing with large or remote deployment areas.

To better understand the complexity of this problem, we can break it down into the following key challenges:

\begin{itemize}
    \item \textbf{Weak Signals and Unstable Connections:} Traditional node deployment methods often result in weak signal strength or unstable connections, particularly in areas with complex terrain or physical obstructions. This leads to frequent data losses and degraded network performance.
    
    \item \textbf{Manual Configuration for Unknown Sites:} Current deployment strategies largely rely on manual visits to configure network parameters for known sites. For unknown or remote locations, this becomes impractical due to the high cost and time required for on-site adjustments.
    
    \item \textbf{Static vs. Mobile Nodes:} While most existing studies focus on optimizing network parameters for static nodes, the requirements of mobile nodes are often overlooked. Mobile nodes introduce additional complexities, such as the need for dynamic reconfiguration of parameters to maintain connectivity while moving through different environments with varying signal conditions.
\end{itemize}

The optimal configuration of LoRaWAN parameters is a non-trivial task that requires careful consideration of the deployment environment. As our study shows, automating the process of configuring these parameters, especially for unknown sites and mobile nodes, is crucial to ensuring the long-term viability of large-scale LoRaWAN deployments.

\subsection{Challenges}

To address the problems outlined earlier, we propose an approach that configures the LoRaWAN network parameters prior to deploying devices at the target site. Before delving into the details of the proposed solution, it is crucial to highlight the key challenges we encountered during the study:

\begin{itemize}
    \item \textbf{Variation in Signal Strength Across Locations:} One of the primary challenges is the variation in signal strength across different locations, even when the distance between the transmitter and receiver remains constant. The signal strength, measured as Received Signal Strength Indicator (RSSI), can fluctuate significantly due to factors such as transceiver height, environmental obstructions, and terrain. For instance, as shown in Figure~\ref{fig:loc}, the RSSI varies when the height of the transmitter changes. This observation underscores the critical role that transceiver height and environmental factors play in determining the quality of the LoRa signal. To mitigate these variations, there is a pressing need to develop a more accurate propagation loss model that takes into account these diverse influences.

    \begin{figure}
        \centering
        \includegraphics[width=1\linewidth]{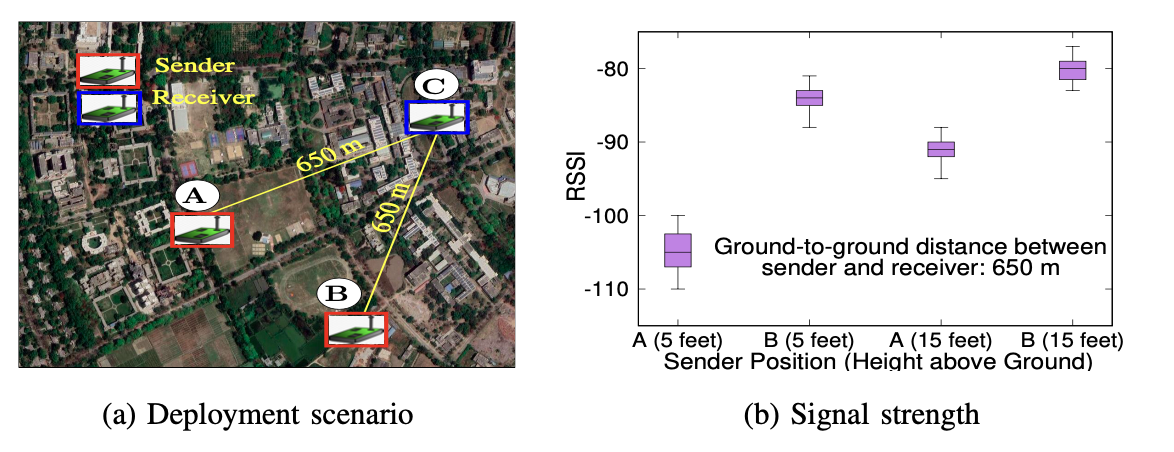}
        \caption{Illustration of signal strength variations across different locations~\cite{conf/networking/PandeyKGRR24}}
        \label{fig:loc}
    \end{figure}

    \item \textbf{Impact of Obstructions and Line-of-Sight (LoS):} Another significant challenge is the impact of physical obstructions on signal strength. While line-of-sight (LoS) communication is often ideal, it is rarely achievable in real-world environments where buildings, trees, and other structures obstruct the signal path. In addition to these direct obstructions, the presence of nearby structures can cause signal reflection, diffraction, or scattering, further degrading signal quality. Figure~\ref{fig:los} compares different LoS settings, highlighting how signal strength weakens in the presence of obstacles. This emphasizes the need to consider both direct obstructions and their surroundings when designing the network, as well as the importance of accurate path loss estimation models that account for these factors.

    \begin{figure}
        \centering
        \includegraphics[width=1\linewidth]{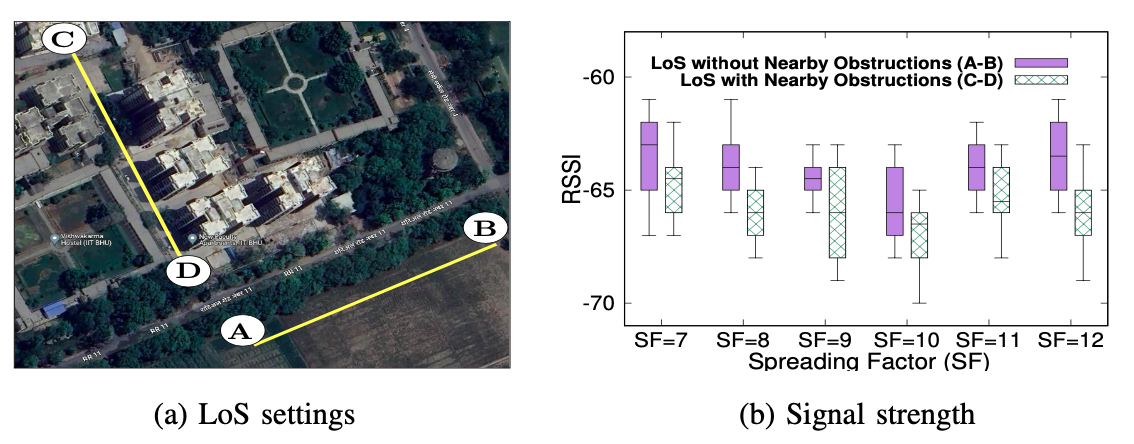}
        \caption{Comparisons between different LoS settings and their impact on signal strength~\cite{conf/networking/PandeyKGRR24}}
        \label{fig:los}
    \end{figure}

    \item \textbf{Challenges of Mobility:} The final challenge relates to the mobility of devices within the LoRaWAN network. Many applications involve mobile nodes that traverse different areas, each with varying environmental conditions. The dynamic nature of such environments introduces variability in the signal strength as the mobile transmitter moves through different regions with diverse obstructions, terrains, and signal interference. Figure~\ref{fig:speed} demonstrates the effect of varying sender speeds on signal strength. These findings highlight the need to consider mobility-related factors, such as the speed of moving devices and the changing environmental conditions, when assessing and optimizing LoRa signal performance. Developing solutions that can dynamically adjust to such changes is crucial for ensuring robust communication in real-world deployments.

    \begin{figure}
        \centering
        \includegraphics[width=1\linewidth]{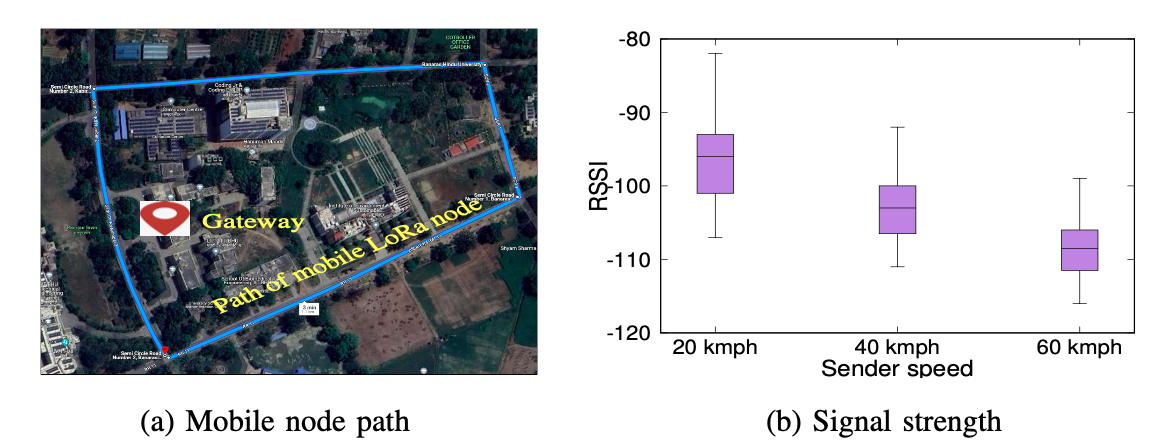}
        \caption{The impact of sender speed on signal strength in mobile environments~\cite{conf/networking/PandeyKGRR24}}
        \label{fig:speed}
    \end{figure}
\end{itemize}

\subsection{Solution Overview}

To address the challenges of configuring LoRaWAN network parameters without requiring on-site visits, we propose a comprehensive approach that combines experimental data, advanced path loss modeling, and automation. This approach eliminates the need for manual network parameter configuration by estimating key variables using environmental data and optimizing the network for specific site conditions. The solution is outlined in the following steps:

\begin{itemize}
    \item \textbf{Multi-Loss Propagation Model:} 
    The core of our solution is the development of a multi-loss propagation model that accurately estimates the path loss between the sender and receiver in a given environment. This model takes into account several factors:
    \begin{itemize}
        \item \textbf{Free Space Loss:} The loss encountered when signals travel in an unobstructed path.
        \item \textbf{Terrain Loss:} Variations in the landscape, such as hills or valleys, which impact signal propagation.
        \item \textbf{Multi-Wall and Floor Loss:} The attenuation caused by walls, floors, and other structural obstacles.
        \item \textbf{Shadowing Effects:} Environmental blockages such as trees, buildings, and other objects that can cause signal shadowing.
    \end{itemize}
    
    This model was developed using data collected from an experimental study conducted on the campus. By measuring the signal strength at different points under various conditions, we created a dataset that feeds into the model. For free space and terrain losses, we used the Ericsson model, while for obstruction losses, we utilized the multi-wall and floor model. Additionally, the First Fresnel Zone (FFZ) in LoRa communication was considered for accurate signal behavior modeling.

    \item \textbf{Obstruction Detection and Path Loss Estimation:} 
    To further enhance the accuracy of our model, it is essential to identify and estimate obstructions along the communication path. MPLoRa introduces an automated method for detecting obstructions using site imagery, thus eliminating the need for manual site surveys. The steps involved include:
    \begin{enumerate}
        \item \textbf{Satellite Image Processing:} We begin by obtaining satellite images of the deployment site. 
        \item \textbf{Shadow Masking and Preprocessing:} A shadow mask is applied to identify areas of obstruction, followed by preprocessing to clean and enhance the image data.
        \item \textbf{Vegetation and Building Segmentation:} The image is further segmented to distinguish between natural obstructions (e.g., trees) and man-made structures (e.g., buildings).
        \item \textbf{Height Estimation:} The heights of buildings and trees are estimated using a combination of satellite image data and 3D modeling. Figure~\ref{fig:height} illustrates the workflow for obstruction detection and height estimation.
    \end{enumerate}
    This automated approach reduces the time and cost associated with manual measurements while providing a high level of accuracy.

    \begin{figure}
        \centering
        \includegraphics[width=1\linewidth]{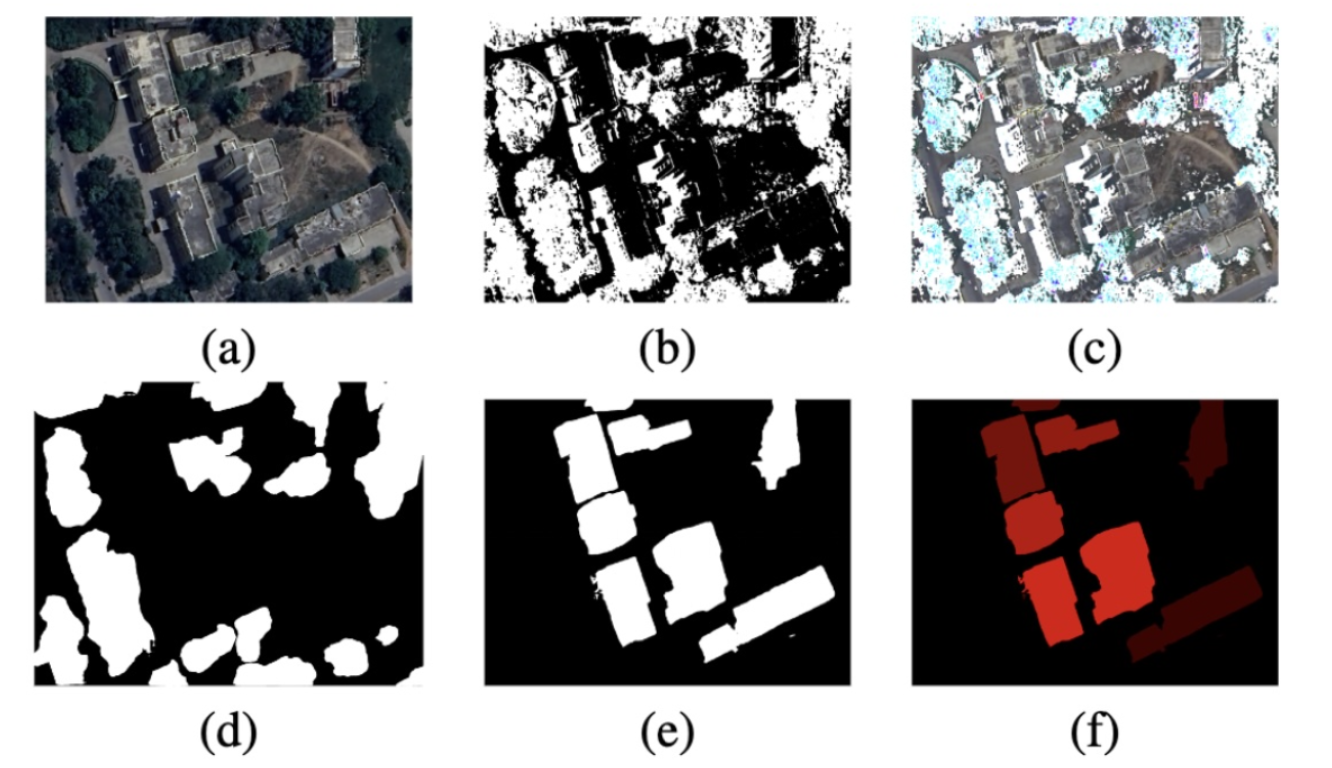}
        \caption{Estimation of height of the obstructions~\cite{conf/networking/PandeyKGRR24}.}
        \label{fig:height}
    \end{figure}

    \item \textbf{Dynamic Network Parameter Adjustment:} 
    After estimating the path loss, MPLoRa dynamically determines the site-specific optimal LoRaWAN parameters, such as the spreading factor (SF) and transmission power (TxPower). Several dynamic environmental factors are considered in this process, including:
    \begin{itemize}
        \item \textbf{Mobility of Transceivers:} The model accounts for the movement of transceivers, which can cause path loss to increase with higher speeds.
        \item \textbf{Campus Infrastructure Changes:} Any long-term changes in the physical layout of the campus (e.g., new buildings) are factored into the recalibration of model variables.
        \item \textbf{Weather and Seasonal Variations:} Fluctuations in weather conditions such as temperature, humidity, and precipitation are considered, as they can significantly affect signal propagation.
    \end{itemize}
    The model continuously recalibrates based on these dynamic factors, ensuring that the network remains optimized for changing environmental conditions.

    \item \textbf{Network Parameter Optimization:}
    The final step in the process is the optimization of LoRaWAN parameters using the estimated path loss from the multi-loss propagation model. The goal is to determine the optimal values of spreading factor (SF) and transmission power (TxPower) that will achieve the desired signal strength while minimizing energy consumption and ensuring reliable communication. The minimum achievable RSSI ($RSSI_{\text{min}}$) for the sender is calculated as:

    \begin{equation}
    RSSI_{\text{min}} = TxP_{\text{min}} + g_s + g_r - P
    \end{equation}

    where:
    \begin{itemize}
        \item $TxP_{\text{min}}$ is the sender’s minimum transmission power,
        \item $g_s$ and $g_r$ are the antenna gains for the sender and receiver, respectively,
        \item $P$ is the path loss.
    \end{itemize}

    The process involves iterating over different values of SF and TxPower, adjusting them to ensure that the calculated RSSI meets the required threshold. Through this optimization, the model determines the most efficient configuration for the LoRaWAN network, balancing signal quality, energy efficiency, and coverage range.
\end{itemize}

\subsection{Results Highlights: Impact of Mobility}

One of the critical aspects investigated in our study is the impact of mobility on LoRaWAN network performance. LoRa is generally used in static or semi-static environments, but as mobile nodes become more common, especially in smart campus applications, it is crucial to understand how mobility affects signal strength, network coverage, and parameter configuration. In this section, we present the key findings regarding the path profile of mobile LoRa nodes and the distribution of spreading factor (SF) and transmission power (TxPower) across different grid points of the deployment site.

\begin{figure}
    \centering
    \includegraphics[width=1\linewidth]{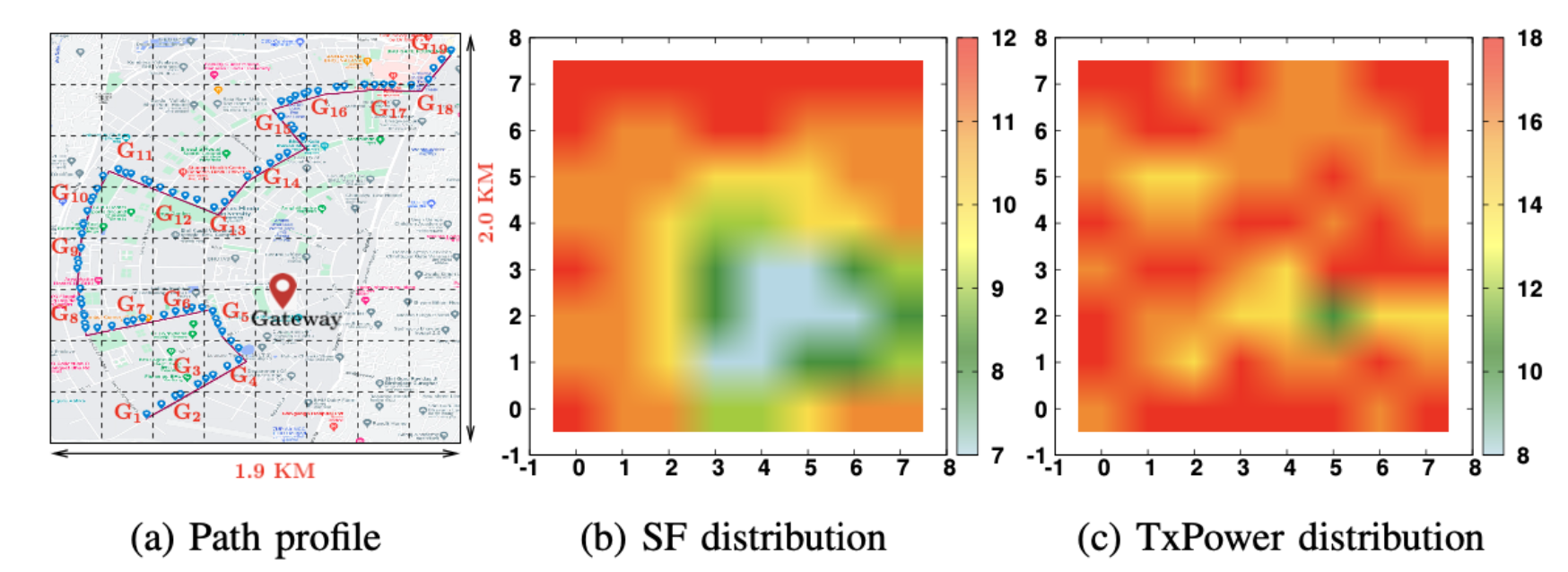}
    \caption{Impact of mobility on LoRa signal strength and performance~\cite{conf/networking/PandeyKGRR24}.}
    \label{fig:moblil}
\end{figure}

\subsubsection{Path Profile of Mobile LoRa Nodes}

The movement of a LoRa node introduces dynamic changes to the communication path between the transmitter and receiver, resulting in fluctuating signal quality and varying propagation losses. In our experiments, we analyzed the mobility of LoRa nodes using pre-determined movement paths across the campus. These paths were selected to cover a range of environments, from open spaces to dense areas with multiple obstructions, simulating real-world movement patterns such as those of vehicles or individuals carrying the nodes.

As illustrated in Figure~\ref{fig:moblil}, the mobile LoRa nodes experience varying degrees of path loss depending on their movement through different regions. We observed that:
\begin{itemize}
    \item In unobstructed, open areas such as courtyards or athletic fields, the path loss remained relatively stable, with minimal variations in signal strength. These regions allowed for maximum signal reach and required lower transmission power (TxPower) to maintain optimal communication.
    \item In areas with buildings, walls, and other obstacles, the path loss increased significantly, especially when nodes moved closer to or behind obstructions. The rapid fluctuation of signal strength in such regions necessitated a higher TxPower to maintain connectivity.
    \item Movement across different elevations, such as walking up and down staircases or moving between buildings of varying heights, caused additional degradation in signal quality. The height difference between the sender and receiver played a key role in the performance of the link, confirming that transceiver height is a significant factor in LoRa signal propagation, as previously discussed in the multi-loss propagation model.
\end{itemize}

These findings underline the importance of continuous path loss estimation in mobile scenarios. Without proper adjustments, a mobile node could easily experience poor connectivity or signal drops in challenging environments, negatively impacting the reliability of the network.

\subsubsection{Distribution of Spreading Factor (SF) and TxPower Across Grids}

In the next phase of our experiment, we divided the deployment area into a grid system, with each grid point representing a specific region of the campus. For each grid, we evaluated the optimal spreading factor (SF) and transmission power (TxPower) necessary to maintain a reliable connection as the mobile node passed through. Figure~\ref{fig:moblil} showcases the distribution of SF and TxPower across these grid points.

\paragraph{Spreading Factor (SF) Variations:}
LoRaWAN’s spreading factor (SF) controls the data rate and communication range. A higher SF increases the range but at the cost of slower data transmission and higher energy consumption. Our study revealed that:
\begin{itemize}
    \item In open spaces where the path loss is minimal, a lower SF (e.g., SF7) was sufficient to maintain communication, allowing for faster data transmission and reduced energy consumption.
    \item In areas with significant obstructions or terrain variations, the SF had to be increased (e.g., SF10 or SF12) to maintain the link quality over longer distances or through walls. While this provided better range, it also introduced delays in data transmission and increased power consumption, particularly for mobile nodes that continuously traverse these regions.
    \item We observed that when the node moved between open and obstructed regions, an adaptive SF strategy was necessary. Rigid configurations (either too high or too low SF) resulted in inefficient energy use or reduced connectivity. Adaptive mechanisms that adjust the SF based on real-time signal conditions proved to be the most effective.
\end{itemize}

\paragraph{TxPower Adjustments:}
The transmission power (TxPower) of a LoRa node determines how far the signal can travel and how much energy is consumed. Our experiments revealed that:
\begin{itemize}
    \item In areas with strong signal reception, such as open fields or when the node was within line of sight (LoS) of the gateway, the required TxPower was minimal. This allowed for significant energy savings, making the LoRaWAN network highly efficient in these regions.
    \item In contrast, areas with obstructions or those located farther from the gateway required a higher TxPower to compensate for signal degradation. This was especially true in multi-wall environments and where the node was mobile and traversing between buildings.
    \item Optimizing TxPower is a delicate balance between maintaining reliable communication and conserving energy. Overestimating the required TxPower in strong signal areas led to unnecessary energy depletion, while underestimating it in weak signal areas resulted in dropped packets and communication failures. The dynamic nature of the mobile node’s path meant that real-time TxPower adjustments were necessary to optimize performance.
\end{itemize}

\subsubsection{Effect of Mobility on Network Performance}

Our study also highlighted the broader impact of mobility on overall network performance. The mobile LoRa node's continuous movement through varied environments led to the following insights:
\begin{itemize}
    \item As the node moved, signal strength fluctuated, with frequent drops and spikes in RSSI (Received Signal Strength Indicator) values. This variability made it difficult to maintain consistent connectivity, particularly in densely populated or obstructed regions.
    \item The variability in signal strength affected data transmission rates. When the node entered regions with higher path loss, the transmission rate had to be reduced, leading to delays. Conversely, in areas with lower path loss, higher transmission rates could be sustained.
    \item The mobile node's power consumption varied significantly across different regions of the deployment site. In areas requiring higher SF and TxPower, the energy usage spiked, particularly during periods of rapid movement. Conversely, in open areas, power consumption dropped, allowing the node to conserve energy effectively.
\end{itemize}

The results demonstrate the need for adaptive mechanisms that continuously monitor and adjust LoRaWAN parameters such as SF and TxPower in response to changes in mobility, environmental conditions, and network requirements. As mobile nodes become more prevalent in smart campus applications, addressing the challenges posed by mobility will be essential for ensuring reliable, efficient, and scalable LoRaWAN networks.

\bibliographystyle{IEEEtran}
\bibliography{paper.bib}

\end{document}